\documentclass{emulateapj}
\usepackage{apjfonts}

\tightenlines

\shorttitle{Effective temperature of galactic O Stars}
\shortauthors{Morisset, C.}

\newcommand{\mysp}{}\def\mysp/{}

\newcommand{\forb}[3]{[\ion{#1\mysp/}{#2}]\-#3$\mu$m}

\newcommand{\hi}{}\def\hi/{\ion{H\mysp/}{1}}
\newcommand{\hii}{}\def\hii/{\ion{H\mysp/}{2}}
\newcommand{\teff}{}\def\teff/{$T_{\rm{eff}}$}
\newcommand{\um}{}\def\um/{$\bar{U}$}
\newcommand{\logg}{}\def\logg/{$\log(g)$}
\newcommand{\logum}{}\def\logum/{$\log(\bar{U})$}
\newcommand{\abne}{}\def\abne/{Ne/Ne$_\odot$}
\newcommand{\abz}{}\def\abz/{$Z/Z_\odot$}
\newcommand{\uchii}{}\def\uchii/{UCHII}
\newcommand{\rgal}{}\def\rgal/{R$_{\rm gal}$}
\newcommand{\exar}{}\def\exar/{[\ion{Ar}{3}/\ion{$\!$}{2}]}
\newcommand{\exs}{}\def\exs/{[\ion{S}{4}/\ion{$\!$}{3}]}
\newcommand{\exne}{}\def\exne/{[\ion{Ne}{3}/\ion{$\!$}{2}]}
\newcommand{\etasne}{}\def\etasne/{$\eta_{\rm S-Ne}$}
\newcommand{\heish}{}\def\heish/{\ion{He\mysp/}{i}/H}
\newcommand{\wmbasic}{}\def\wmbasic/{{\it WM-Basic}}
\newcommand{\costar}{}\def\costar/{{\it CoStar}}
\newcommand{\tlusty}{}\def\tlusty/{{\it TLUSTY}}
\newcommand{\cmfgen}{}\def\cmfgen/{{\it CMFGEN}}
\newcommand{\kurucz}{}\def\kurucz/{{\it Kurucz}}
\newcommand{\msbm}{}\def\msbm/{MSBM03}
\newcommand{\figdia}{}\def\figdia/{Fig.~\ref{fig:all}}

\begin{document}

\title{O stars effective temperature and \hii/ regions ionization parameter
\\ gradients in the Galaxy}

\author{C. Morisset}
\affil{Instituto de Astronom\'{\i}a, Universidad Nacional
        Aut\'onoma de M\'exico; Apdo. postal 70--264; Ciudad Universitaria;
        M\'exico D.F. 04510; M\'exico. {\it morisset@astroscu.unam.mx} 
	}

\begin{abstract}
Extensive photoionization model grids are computed for single star
\hii/ regions using stellar atmosphere models from the \wmbasic/ code.
Mid-IR emission line intensities are predicted and diagnostic diagrams
of \exne/ and \exs/ excitation ratio 
are build, taking into account the metallicities
of both the star and the \hii/ region. The diagrams are used in
conjunction with galactic \hii/ region observations obtained with the ISO
Observatory to determine the effective temperature \teff/ of the
exciting O stars and the   
mean ionization parameter \um/. \teff/ and \um/ are found to
increase and decrease, respectively, with the metallicity of the
\hii/ region represented by the \abne/ ratio. No evidence is found for
gradients of \teff/ or \um/ with galactocentric distance \rgal/.
The observed excitation sequence with \rgal/ is mainly due to the
effect of the metallicity gradient on the spectral ionizing shape,
upon which the effect of an increase in \teff/ with $Z$ is
superimposed. We show that 
not taking properly into account the effect of metallicity on the
ionizing shape of the stellar atmosphere would lead to an apparent
decrease of \teff/ with $Z$ and an increase of \teff/ with \rgal/. 
\end{abstract}


%------------------------------------------------------------------
\section{Introduction}
%------------------------------------------------------------------

The determination of the stellar distribution (especially of the
hottest stars) and physical characteristics of 
galactic \hii/ regions are of primary importance to evaluate star
formation theories and for our understanding of the chemical evolution
of galaxies. 

\citet{ST76} used the equivalent width of the H$_\beta$ emission from
\hii/ regions in spiral galaxies to determine the existence of a radial
gradient in the 
effective temperature (hereafter \teff/) of the hottest stars,
associated with a 
decrease in metal abundance. \citet{C88} determined \teff/
and the ionization parameter $U$ for various \hii/ galaxies, and
concluded that the \teff/ of 
the hottest star decreases with increasing oxygen abundance.

On the other hand, \citet{ED85} have computed extensive photoionization
models, using 
\citet{HM70} atmosphere models, and have determined from optical observations
of \hii/ regions that the ionization temperature of the exciting stars
is approximately constant (41~kK, independently of the metallicity $Z$
and $U$). They found, however, an anticorrelation between
$U$ and $Z$. \citet{FTP86} found a near constant \teff/ of 35~kK 
between 1 and 5~kpc from the center of NGC~2403. 

More recently, \citet{MVT02} used Infrared Space Observatory (ISO)
spectral observations of galactic \hii/ 
regions to show that the gas excitation increases with the
galactocentric distance \rgal/. They concluded that the stellar spectral
energy distributions (hereafter SEDs) are
softer at higher metallicities, that is towards the galactic
center and that the SED changes can explain the observed gradient.
\citet{GSL02} similarly used ISO observations, but suggest 
that the increase in excitation correspond instead to a
decrease in stellar effective temperature. 
\citet[][hereafter \msbm/]{MSBM03} show that excitation gradients
are partly due to changes in the ionizing 
spectral shape of O stars with metallicity. They concluded that
the excitation scatter is probably mainly due to randomization 
of both the stellar \teff/ and the nebular mean ionization parameter
\um/\footnote{The mean ionization parameter \um/ is defined 
following \citet{ED85} as the value of $U$ evaluated at a distance
from the ionizing star $\bar{r} = 
r_{empty}$ + $\Delta$R/2, where $r_{empty}$ is the size of the empty
cavity and $\Delta$R is the thickness of the uniform density \hii/ shell.}.

No attempt was made by \citet{GSL02}, \citet{MVT02}, nor in \msbm/
to determine \teff/ and $U$ for individual \hii/ regions.

\citet{DP03} used optical observations of galactic and magellanic
\hii/ regions to determine \teff/ from optical diagnostic line
ratios. They also found an increase of \teff/ with the galactocentric
distance.

The aim of the present work firstly is to build diagnostics
diagrams for the determination of \teff/ and \um/, based on mid-IR
emission lines. The diagrams are derived from a extensive grid of
photoionization models that populate the \teff/-\um/-$Z$ space and use
the \wmbasic/ \citep{PHL01} code to compute the ionizing atmosphere models.
In a second step, \teff/ and \um/ are determined for the ISO \hii/
regions using the new diagnostic diagrams.

Sect.~\ref{sec:obs} describes the ISO observations of \hii/ regions, 
and Sect.~\ref{sec:models} the grid of photoionization models.
The location of ISO observations in the model
grids, and the process to determine \teff/, and the mean ionization
parameter \um/ for every
object are presented in Sect.~\ref{sec:teffu}, using two different methods.
Sect.~\ref{sec:results} describes the resulting gradients of \teff/
and \um/. The effect of the stellar metallicity in
the determination of \teff/ is discussed in Sect.~\ref{sec:disc}, in
particular the influence of the changes in the stellar SEDs with
metallicity. The conclusions are presented in Sect.~\ref{sec:conc}.

%------------------------------------------------------------------
\section{ISO Observations of \hii/ regions}
\label{sec:obs}
%------------------------------------------------------------------

Mid-IR fine-structure line intensities obtained from observations of
\hii/ regions with ISO-SWS \citep{PaperII,GSL02} are 
used here to determine the various stellar and nebular
parameters. 

Line ratios of 
\forb{Ar}{3}{8.98} / \forb{Ar}{2}{6.98},
\forb{S}{4}{10.5} / \forb{S}{3}{18.7}, and
\forb{Ne}{3}{15.5} / \forb{Ne}{2}{12.8} (hereafter \exar/, \exs/ and
\exne/ resp.) are then used to build excitation diagrams (see \msbm/ for
more details). The line intensities were corrected for
reddening by \citet{GMS02}. Once the sources for which at least one of
the line intensities used in this work is not defined (or have only an
upper limit) are removed, 42 usable sources remain.

%\newpage
%\vspace{1.cm}
%------------------------------------------------------------------
\section{Grid of photoionization models}
\label{sec:models}
%------------------------------------------------------------------

A grid of photoionization models using the NEBU code \citep{PaperIII}
has been calculated using ionizing spectral 
distributions from supergiant atmosphere models that were computed with
\wmbasic/~V.~2.11\footnote{freeware available
at http://www.usm.uni-muenchen.de/people/adi/} \citep{PHL01}. 
The current grid of photoionisation models is similar to the one used
in \msbm/ (see \msbm/ for details). It has been extended further to
encompass the full range of values expected within the 3D
parameters space \teff/-\um/-$Z$. \teff/ is ranging from 30 to 50~kK,
by 5~kK steps, 
\logum/ being -2.6, -1.5, -0.8, -0.1 and 0.5, and the metallicity $Z$
of both the 
ionizing star and the nebular gas being 0.5, 0.75,
1.0, 1.5 and 2.0 times the solar value (as defined in \wmbasic/). 
In total, 125 models have been computed from which mid-IR line
intensities were derived.

%

%------------------------------------------------------------------
\section{Determination of \teff/ and \um/}
\label{sec:teffu}
%------------------------------------------------------------------
Three mid-IR line ratios could in principle be used as excitation diagnostics, 
namely \exar/, \exne/ and \exs/). We note however that \exar/ is
overestimated in photoionization models, as pointed out in \msbm/. It is not
clear whether the latter is due to an improper determination of the
ionizing flux near 24~eV or simply to incomplete atomic physics used within 
photoionization codes (e.g. missing accurate dielectronic recombination rates). 
For these reasons, despite its low dependence on \um/, the \exar/
ratio proves to be useless for determining \teff/.
On the other hand, while the alternative excitation ratios 
\exne/ and \exs/ are both sensitive to \teff/
and \um/, it turns out that the effect of either parameter on both
ratios is somewhat 
different, and a way to determine \teff/ and \um/ using these
excitation diagnostics can be extracted. 
Even though the direct use of \exne/ and \exs/ or other combinations of
these line ratios would provide equivalent constraints; we prefer to adopt 
\etasne/, defined following \citet{VP88} as \etasne/ = \exs/ /
\exne/, in combination with \exne/. 
The determination of \teff/ and \um/ turns out to be clearer and easier
to read off when using \etasne/.

The underlying hypothesis/assumptions
made here are the following: 
1) the ISO \hii/ regions are excited by a single star (i.e. the
presence of other less luminous stars doesn't affect the results), 
2) the \hii/ regions are ionized by stars of comparable surface
gravity \logg/ at a given \teff/ (see \msbm/ for the effects of \logg/),
3) the Ne abundance determined from an \hii/ region is a reliable
estimator of the ionizing star metallicity, 
4) the presence of dust in the \hii/ regions doesn't affect the
\teff/ and \um/ gradients. Dust would decrease the global amount of
ionizing photons, but also increase the hardness of the ionizing SED,
increasing the excitation of the gas depicted by the IR excitation
diagnostics (see \msbm/) and therefore the \teff/ we infer can be
overestimated,
5) Using a nebular geometry consisting of a simple shell does not
affect the diagnostic diagrams,
6) the \wmbasic/ atmosphere models describe well the ionizing flux
between 35 and 41~eV (or at least the relative changes that occurs
when the parameters \teff/ or $Z$ are varied). 

%------------------------------------------------------------------
\subsection{S-Method : Using only solar metallicity atmosphere models}
%------------------------------------------------------------------
\label{sec:met1}
In a first step, we use only the results of the photoionization
models obtained with the solar abundances atmosphere models. 

\begin{figure}
\epsscale{1.0}
\plotone{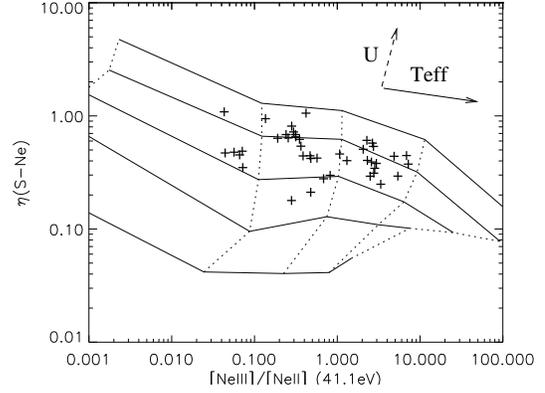}
\caption{The \etasne/ versus \exne/ excitation diagram for the \hii/ regions
observed by ISO.
The grid of photoionization model with stellar and nebular
metallicity given by \abz/ = 1.0 (S-Method) is drawn. Solid (dotted) lines 
connect iso-\um/ (iso-\teff/) models. \teff/ and \logum/ take
values of 30, 35, 40, 45, 50~kK and -2.6, -1.5, -0.8, -0.1, 0.5,
respectively.
 \label{fig:diag_sol}}
\end{figure}

The Fig.~\ref{fig:diag_sol} shows the values taken by \etasne/ and \exne/, when
\teff/ and \um/ are varied in locked steps. For each \hii/ region,
2D-interpolations within this grid are performed and used to determine
\teff/ and \um/. All the observed values lie inside the grids, no
extrapolation is needed. The
\teff/ and \um/ obtained with solar metallicity atmosphere models are
presented in Sec.~\ref{sec:results}.

From Fig.~\ref{fig:diag_sol}, we can determine the effects of
uncertainties in line intensities on \teff/ and\um/: a factor of two
in the excitation diagnostics leads to a shift in \teff/ by 1~kK and
in \um/ by 0.5~dex.

%------------------------------------------------------------------
\subsection{Z-Method : \teff/ and \um/ obtained using Z-dependent atmosphere models}
%------------------------------------------------------------------
\label{sec:met2}
The chemical composition of a star strongly affects its radiation,
especially in the EUV, where the ionizing photons are emitted. 
For the same \teff/, changing the stellar luminosity by a factor of 4
can affect the excitation diagnostic line ratios by up to 2 orders of
magnitude (\msbm/). The determination of \teff/ and \um/ need then to
be performed using grids of photoionization models with various
stellar metallicities, as described in Sect.~\ref{sec:models}.
Making
the assumption that the metallicity of the ionized region reflects the
metallicity of the ionizing star, we can use adapted diagnostic
grids, corresponding to the \hii/ regions metallicities, to determine \teff/
and \um/.
The metallicity of an \hii/ region is hereafter given by the abundance
ratio \abne/ = [Ne/H]/[Ne/H]$_\odot$, where [Ne/H] is obtained from
\citet{GMS02}. The solar abundance used, [Ne/H]$_\odot = 1.4\times
10^{-4}$, is defined as the value of the abundance gradient [Ne/H](\rgal/) found
by \citet{GMS02}, evaluated at 8.5~kpc. 
The Ne abundances as a measure of \abz/, are preferred to a combination
of Ne, Ar and S abundances, since,  
for these two last elements, the abundances are not reliable when
the excitation is extreme (\msbm/). 
For each \hii/ region, we extract from the \teff/-\um/-$Z$
photoionization model grid the \teff/-\um/ plane, whose
metallicity lies closest to the \abne/ of the \hii/ region. 
The \figdia/ shows, for the 5 metallicities used in the photoionization
model grid, the values taken by \etasne/ and \exne/. 
The effect of increasing the metallicity of the
atmosphere models (and consequently of the nebular gas) is
clearly to increase the value of \teff/ for a given \exne/
ratio, as already pointed out in \msbm/.
The observed values for the \hii/ regions 
are also plotted in the diagramm corresponding to their metallicities. 
All the observed values lie inside the grids. 2D-Interpolations are
performed to determine \teff/ and \um/ for each \hii/ region.

\begin{figure*}
\epsscale{0.95}
\plotone{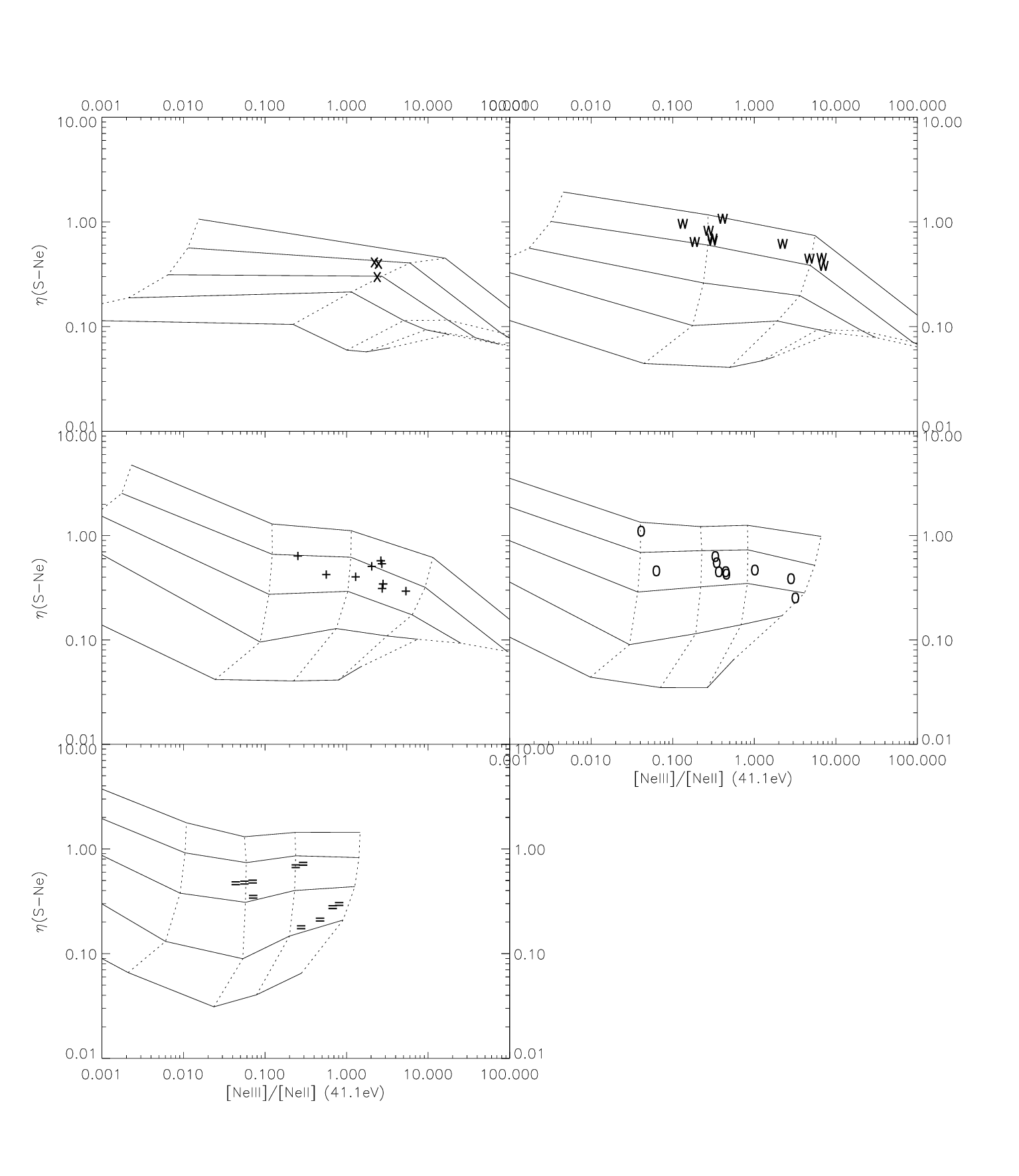}
\caption{Same as Fig.~\ref{fig:diag_sol} but the stellar and nebular
metallicities beeing  0.5,
0.75, 1.0, 1.5, 2.0 times solar, from upper left to lower right. Solid
(dotted) lines connect iso-\um/ (iso-\teff/) models. \teff/ and
\logum/ take values of 30, 35, 40, 45, 50~kK and -2.6, -1.5, -0.8,
-0.1, 0.5, respectively.
All the observations of the \hii/ regions are distributed in the diagramms,
according to their nearest values of \abne/ (Z-Method). 
The \hii/ regions are symbolized using X, W, +, 0, and = for
\abne/ = 0.5, 0.75, 1.0, 1.5, 2.0, respecively. See
Tab.~\ref{tab:res} for the correspondance between sources and the
diagramm used.
\label{fig:all}}
\end{figure*}

%------------------------------------------------------------------
\section{Results}
\label{sec:results}
%------------------------------------------------------------------

The Table~\ref{tab:res} presents the characteristics of the 42 \hii/
regions used in this work: \rgal/, \abne/ (with a symbol corresponding
to the grid used within the Z-Method, Sec.~\ref{sec:met2}), \exne/,
\exs/, and the 
\teff/ and \um/ obtained using the Z-Method. The \teff/ range from 34 to  
50~kK, and \logum/ from -1.5 to 0.5, with mean values of 40.5~kK and
-0.60, respectively. Such range for \teff/ and \um/ is in agreement
with the results found by \citet{ED85}. For the 3 sources for which two
independent observations are available, the results obtained lead to a
coherent determination of \teff/, while the values of \um/ can differ
by a factor up to 5. 

\begin{figure*}
\epsscale{0.95}
\plotone{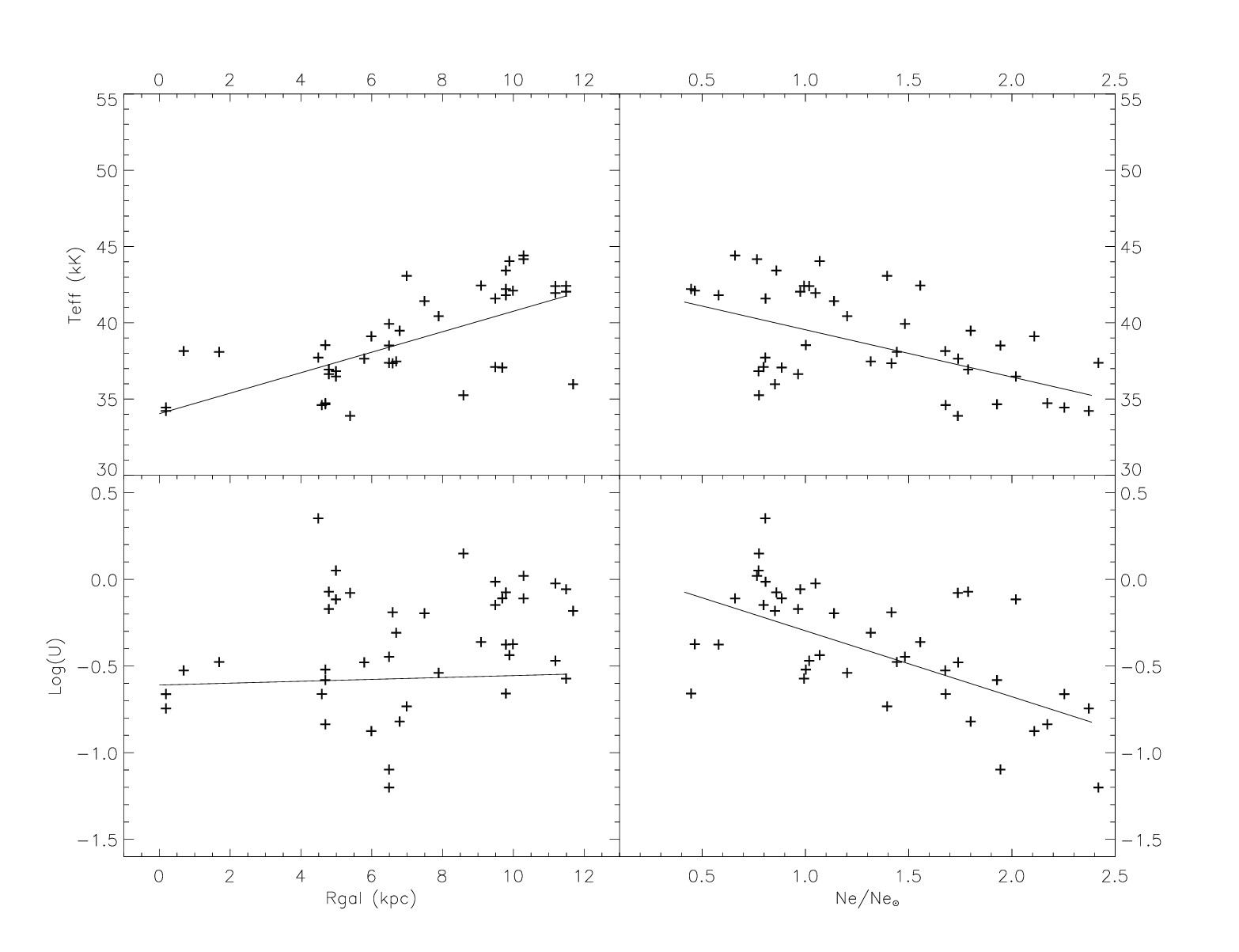}
\caption{\teff/ (upper panels) and \um/ (lower panels) versus \rgal/ (left)
and \abne/ (right) for the sample of \hii/ regions. The grid of
photoionization models results computed with 
$Z=Z_\odot$ stars show in 
Fig.~\ref{fig:diag_sol} is used for every source (S-Method,
Sec.~\ref{sec:met1}). Linear fits to the data are presented.  
\label{fig:res_1z}}
\plotone{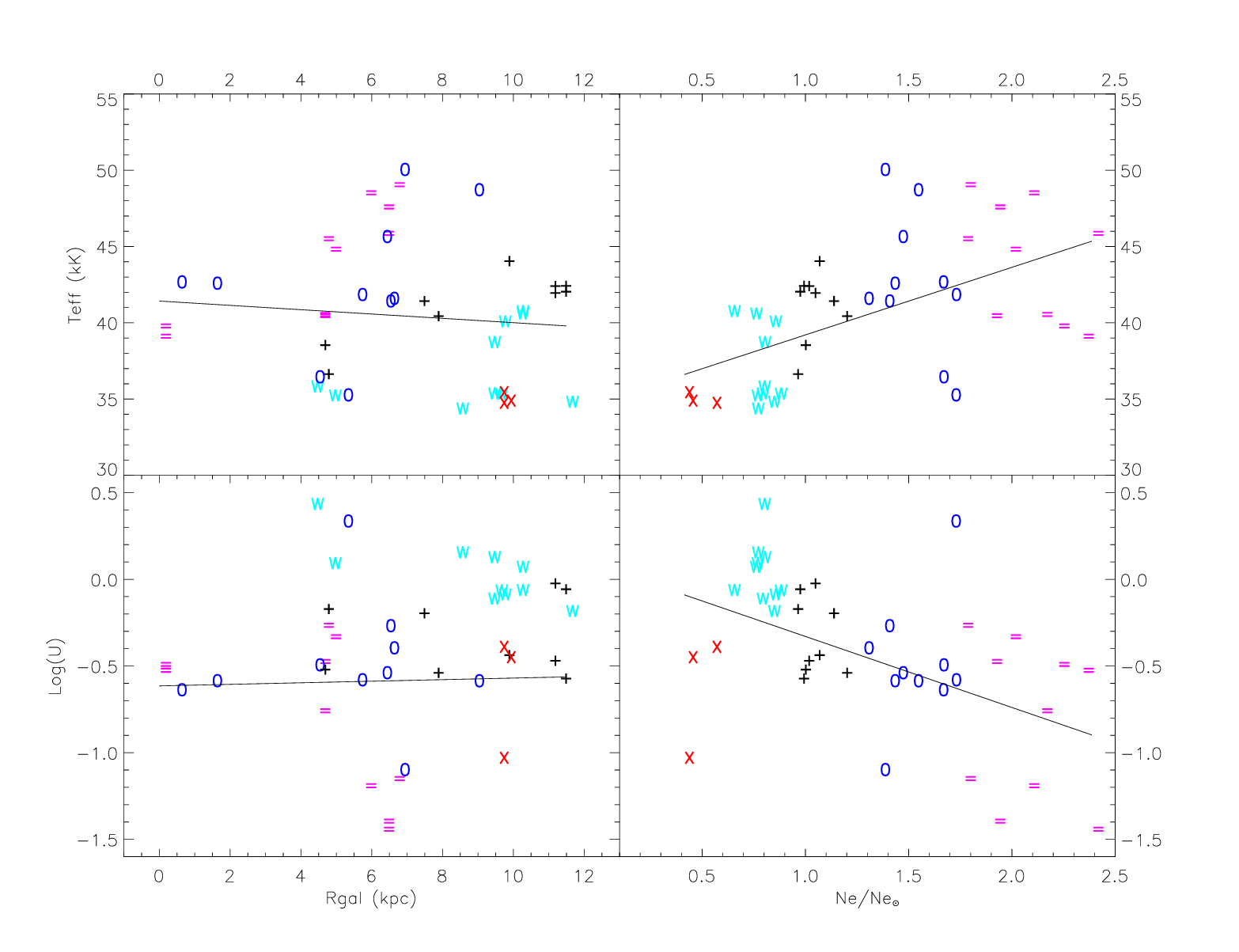}
\caption{Same as in Fig.~\ref{fig:res_1z}, but each \hii/ region
is symbolized according to its abundance \abne/ as in
Fig.~\ref{fig:all}. The determination of
\teff/ and \um/ follows from the use of the appropriate diagnostic
diagram from Fig.~\ref{fig:all} (Z-Method,
Sec.~\ref{sec:met2}). Linear fits to the data are presented. \label{fig:res_all}} 
\end{figure*}

The set of  inferred values for \teff/ and \logum/ versus the galactocentric
radius \rgal/ and the abundance ratio \abne/ are shown in
Fig.~\ref{fig:res_1z} for the S-Method (Sec.~\ref{sec:met1}), and in Fig.~\ref{fig:res_all}
for the Z-Method (Sec.~\ref{sec:met2}). Linear fit to the data are also
presented in all the figures.

The results obtained for \teff/ with S- and Z-Methods are very
different (upper panels of Figs.~\ref{fig:res_1z}-\ref{fig:res_all}).
While S-Method leads to an increase (decrease) of \teff/ with \rgal/
(\abne/), the use of the Z-Method leads to rather different results:
no clear correlation is found to exist between \teff/ and \rgal/ while
a correlation is present between \teff/ and \abne/. The distribution
of \teff/ versus \abne/ obtained with the Z-Method can also be
described as follows: \teff/ increasing
strongly with \abne/ for \abne/$<1.2$, and a quasi constant value
($\sim43$~kK) for higher \abne/, coupled with a higher dispersion.

The dispersion of \abne/ with position in the Galaxy is relatively
high (see the symbols dispersion along \rgal/ in
the left panels of Fig.~\ref{fig:res_all}). The absence of
correlation between \teff/ and \rgal/ obtained with Z-Method might be
the result of this high 
dispersion. Note that for \hii/ regions with
\rgal/$<$\rgal/$_\odot$, the determination of \rgal/ is degenerate
\citep{PaperII} and the errors are also important.

Virtually no changes are observed for \logum/ from
S- to Z-Method (lower panels of Figs.~\ref{fig:res_1z}-\ref{fig:res_all}). 
This gives insights of the robustness of
our results concerning \um/ whatever the method used. 
This can be understood as follows: the main effect of changing $Z$ on the
diagnostic diagrams (\figdia/) is to shift horizontally (\exne/ axis)
the grids of models, while the determination of \um/ is mainly
dependent on the vertical position along the \etasne/ axis, which is
not affected by the stellar metallicity. 
No clear correlation is found between \um/ and \rgal/ but an inverse
correlation between \um/ and \abne/ is present (see
Fig.~\ref{fig:res_all}, lower panels).

\begin{figure}
\epsscale{1.0}
\plotone{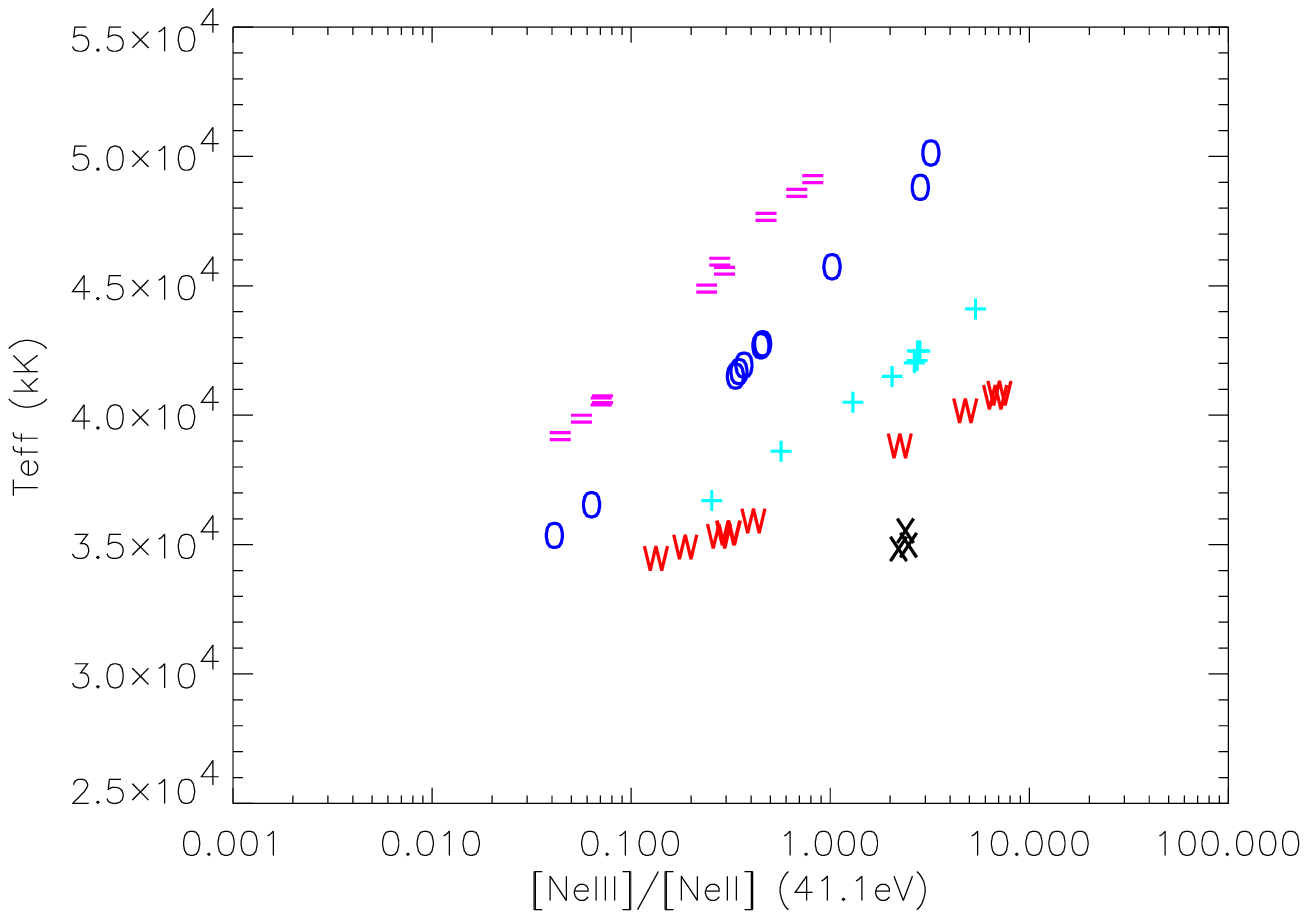}
\caption{\teff/ versus \exne/
for the ISO \hii/ regions. Symbols as in Fig.\ref{fig:res_all}.
\label{fig:teff_exne}}
\end{figure}

Fig.~\ref{fig:teff_exne} shows the distribution of \teff/ versus
the \exne/ excitation ratio. 
The softening of the stellar emission when the
metallicity increases (even if \teff/ doesn't change) is enough to
cover the whole observed range of \exne/ (over 2.5 dex). This
Fig.~\ref{fig:teff_exne} illustrates one more time the 
illusion of determining \teff/ from only one excitation diagnostic.On
the other hand, a global increase of \teff/ with \exne/ is also
present (the hottest stars are associated with the most excited \hii/
regions). 

\begin{table}
\begin{tabular}{lcccccc} 
\tableline
        Name$^a$      &\rgal/$^b$ &\abne/$^c$   &\exne/$^c$  &\exs/$^d$
	&\teff/$^{e,f}$  & \um/$^f$ \\ 
\tableline
 GCRINGSW           & 0.00  &2.221$^{\rm =}$  & 0.048  & 0.021  &39.5  &0.302\\
 ARCHFILNW          & 0.00  &2.340$^{\rm =}$  & 0.037  & 0.016  &38.8  &0.280\\
   IRAS 17455-2800  & 0.50  &1.645$^{\rm O}$  & 0.405  & 0.156  &42.3  &0.213\\
 SGR D HII          & 1.50  &1.410$^{\rm O}$  & 0.398  & 0.161  &42.2  &0.241\\
 WBH9815567-5236  & 4.30  &0.773$^{\rm W}$  & 0.355  & 0.343  &35.5  &2.521\\
 RAFGL 2094         & 4.40  &1.647$^{\rm O}$  & 0.056  & 0.023  &36.1  &0.297\\
   IRAS 18317-0757  & 4.50  &1.895$^{\rm =}$  & 0.060  & 0.027  &40.1  &0.314\\
 WBH9818317-0757  & 4.50  &2.139$^{\rm =}$  & 0.061  & 0.020  &40.2  &0.163\\
 GAL 033.91+00.11   & 4.50  &0.970$^{\rm +}$  & 0.482  & 0.187  &38.2  &0.282\\
   IRAS 15502-5302  & 4.60  &0.932$^{\rm +}$  & 0.216  & 0.126  &36.3  &0.630\\
   IRAS 18434-0242  & 4.60  &1.755$^{\rm =}$  & 0.251  & 0.165  &45.2  &0.508\\
   IRAS 18502+0051  & 4.80  &1.987$^{\rm =}$  & 0.204  & 0.127  &44.5  &0.437\\
 GRS 328.30+00.43   & 4.80  &0.741$^{\rm W}$  & 0.239  & 0.177  &34.9  &1.148\\
   IRAS 17221-3619  & 5.20  &1.706$^{\rm O}$  & 0.036  & 0.036  &34.9  &2.007\\
 WBH9816172-5028  & 5.60  &1.707$^{\rm O}$  & 0.327  & 0.132  &41.5  &0.244\\
 GAL 337.9-00.5     & 5.80  &2.076$^{\rm =}$  & 0.578  & 0.147  &48.2  &0.060\\
 G327.3-0.5         & 6.30  &1.912$^{\rm =}$  & 0.405  & 0.078  &47.3  &0.038\\
 WBH9817059-4132  & 6.30  &2.386$^{\rm =}$  & 0.237  & 0.039  &45.5  &0.034\\
 GAL 045.45+00.06   & 6.30  &1.450$^{\rm O}$  & 0.905  & 0.381  &45.3  &0.268\\
   IRAS 15384-5348  & 6.40  &1.384$^{\rm O}$  & 0.296  & 0.168  &41.0  &0.500\\
 GRS 326.44+00.91   & 6.50  &1.284$^{\rm O}$  & 0.308  & 0.152  &41.2  &0.372\\
 WBH9815408-5356  & 6.60  &1.768$^{\rm =}$  & 0.695  & 0.189  &48.7  &0.066\\
 M17IRAMPOS8        & 6.80  &1.363$^{\rm O}$  & 2.839  & 0.646  &49.7  &0.074\\
 W51 IRS2           & 7.30  &1.106$^{\rm +}$  & 1.743  & 0.807  &41.1  &0.596\\
 NGC6357I           & 7.70  &1.169$^{\rm +}$  & 1.107  & 0.407  &40.1  &0.270\\
 S106 IRS4          & 8.40  &0.742$^{\rm W}$  & 0.115  & 0.099  &34.0  &1.326\\
 NGC3603            & 8.90  &1.524$^{\rm O}$  & 2.513  & 0.873  &48.3  &0.240\\
   IRAS 12063-6259  & 9.30  &0.775$^{\rm W}$  & 1.930  & 1.071  &38.4  &1.242\\
 GAL 289.88-00.79   & 9.30  &0.765$^{\rm W}$  & 0.267  & 0.159  &35.0  &0.716\\
   IRAS 10589-6034  & 9.50  &0.853$^{\rm W}$  & 0.265  & 0.165  &35.0  &0.799\\
 RAFGL 4127         & 9.60  &0.827$^{\rm W}$  & 4.116  & 1.654  &39.7  &0.759\\
 BE83IR 070.29+0  & 9.60  &0.547$^{\rm X}$  & 1.954  & 0.721  &34.4  &0.377\\
 BE83IR 070.29+0  & 9.60  &0.414$^{\rm X}$  & 2.115  & 0.566  &35.1  &0.086\\
   IRAS 11143-6113  & 9.70  &1.037$^{\rm +}$  & 4.571  & 1.226  &43.7  &0.342\\
   IRAS 19598+3324  & 9.80  &0.432$^{\rm X}$  & 2.192  & 0.784  &34.5  &0.329\\
   IRAS 12073-6233  &10.10  &0.733$^{\rm W}$  & 5.813  & 2.368  &40.2  &1.093\\
 GAL 298.23-00.33   &10.10  &0.627$^{\rm W}$  & 6.110  & 2.091  &40.4  &0.803\\
   IRAS 02219+6152  &11.00  &0.986$^{\rm +}$  & 2.402  & 0.755  &42.1  &0.317\\
   IRAS 02219+6152  &11.00  &1.016$^{\rm +}$  & 2.252  & 1.182  &41.6  &0.887\\
 W3 IRS2            &11.30  &0.961$^{\rm +}$  & 2.339  & 0.665  &42.1  &0.250\\
 W3 IRS2            &11.30  &0.943$^{\rm +}$  & 2.321  & 1.139  &41.7  &0.820\\
 S156 A             &11.50  &0.820$^{\rm W}$  & 0.161  & 0.094  &34.4  &0.607\\
\tableline
\end{tabular}
\caption{\label{tab:res} Sources parameters and corresponding \teff/ and \um/
determined in this paper.
$^a$: IRAS named sources from \citet{PaperII},
otherwise from \citet{GSL02}. 
$^b$: \rgal/ in Kpc.
$^c$: Upper symbols indicat which grid is used to determine
\teff/ and \um/: X, W, +, O, = correspond to the \abne/ = 0.5, 0.75,
1.0, 1.5 and 2.0 grid resp. (See Fig.~\ref{fig:all})
$^d$: Line intensities corrected for reddening, see text.
$^e$: \teff/ in kK.
$^f$: Values obtained using the Z-Method (Sec.~\ref{sec:met2}).}
\end{table}

%------------------------------------------------------------------
\section{Discussion}
\label{sec:disc}
%------------------------------------------------------------------

The results presented in Sect.~\ref{sec:results} are sensitive to the
changes of the stellar SED with metallicity, for a
given atmosphere model code. 
The gradients in \teff/ as a function of \abne/ and \rgal/
derived from a single solar metallicity set of stars (S-Method) are very
different from the gradients obtained with coherent
metallicity for the stellar atmosphere model (Z-Method).

The results obtained with the S-Method agree with those of
\citet{C88} and \citet{DP03}, 
who similarly didn't take into account the $Z$-dependence
of the stellar emission.
This confirms that the trends seen in
the upper panels of Fig.~\ref{fig:res_all}, when the $Z$-dependence is
properly considered, are mainly due to the
changes in the stellar SEDs when the 
metallicity decreases (an effect which has to be present, but
whose magnitude might depend
on the family of atmosphere models used). 
\citet{DP03} check the effect of metallicity
on the excitation of the ionized gas, but didn't found a strong
effect; the metallicity range they test is only a factor of two for
\wmbasic/ models, they used a low value for the ionization parameter,
the effect of $Z$ being reduced in this case, and they used dwarf
\wmbasic/ models, for which the 
effect of $Z$ is lower than for supergiants.
The \teff/ values obtained by \citet{C88} and \citet{DP03} are derived using a
S-Method, and the apparent decrease of \teff/ with $Z$ and increase of \teff/ with \rgal/ that
they respectively found is likely to be the result of not 
considering $Z$-dependent 
stellar models (the excitation of the gas is not a valid \teff/
indicator, as shown by Fig.~\ref{fig:teff_exne}). 
Note also that the maximum \teff/ obtained with S-Method is 
45~kK, while the \teff/ values from Z-Method get as
high as 50~kK. 

The results shown in this paper concerning \teff/ and \um/ are
strongly dependent on the kind of atmosphere model used to compute the
photoionization grid of models. In \msbm/, we discuss in detail
the major differences between, for instance, \wmbasic/ and
\cmfgen/ \citep{HM98}, in sofar as 
the determination of \teff/ is concerned. Using the \cmfgen/ 
instead of \wmbasic/ atmosphere models would certainly lead to a
globally lower \teff/ (\msbm/), and perhaps different gradients.

The increase of \teff/ with $Z$ found here is 
in contradiction with the theoretical predictions of
e.g.~\citet{SSMM92}, and if confirmed it might have profond
implications for the study of the upper mass end of the IMF.

However, more extensive studies will be needed to
check whether the use of different atmospheres codes will confirm
the gradients found here or generate genuine gradients.
Our results, nevertheless, show the importance of taking
properly into account the variation in the stellar SEDs with metallicity
in any attempt to determine a reliable \teff/ from \hii/ regions.

%------------------------------------------------------------------
\section{Summary and Conclusion}
\label{sec:conc}
%------------------------------------------------------------------

Based on \wmbasic/ atmosphere models we have computed a large set of
photoionisation models. From these models we have built excitation
diagnostic diagrams based on \exne/ and \exs/ (mid-IR lines)
excitation ratios. 
ISO observations of galactic \hii/ regions are superimposed to
these diagrams. According to their metallicity, \teff/ and \um/
are determined for every \hii/ region. 

A correlation between \teff/ and
\abne/, and an anti-correlation between \um/ and \abne/, have been
found, without evidence of any correlation between both \teff/ and
\um/ versus \rgal/. The determination of \teff/ 
is strongly dependent on the changes in stellar SEDs due to
the radial metallicity gradient within the Galaxy, while the results
found concerning the behaviour of \um/ 
are globally insensitive to this effect. The gaseous excitation
sequence is therefore mainly driven by the effects of metallicity on
the stellar SEDs. 
A global increase of \teff/ with metallicity appears nevertheless to
be present. 
More investigation  using different atmosphere codes
will be needed to confirm that our conclusions are not unduly biased
toward the use of \wmbasic/ models. 
Comparison with \teff/ determined from direct observations of ionizing
stars can also help to evaluate the robustness of the method presented
in this work.

%------------------------------------------------------------------
\acknowledgments
I am gratefull to Luc Binette and Manuel Peimbert for useful
comments and a critical reading of the manuscript. 
This work was partly supported by the CONACyT (M\'exico) grant 40096-F. 
%------------------------------------------------------------------
%\clearpage
%\bibliography{uchii}
%\bibliographystyle{apj}

%------------------------------------------------------------------

\end{document}